\documentclass[aps,prb,twocolumn,superscriptaddress,nofootinbib]{revtex4-1}
\usepackage[colorlinks=true,citecolor=blue,linkcolor=blue,urlcolor=blue]{hyperref}
\usepackage{amsmath, amssymb}
\usepackage{graphicx, nicefrac, textcomp}
\usepackage{color}
\usepackage{dcolumn}
\usepackage[utf8]{inputenc}
\usepackage{float}
\usepackage{xspace,dsfont}

\begin{document}

\title{A superlattice approach to doping infinite-layer nickelates}

\author{R.~A.~Ortiz}
\affiliation{Max Planck Institute for Solid State Research and Center of Integrated Quantum Network, Heisenbergstra{\ss}e 1, 70569 Stuttgart, Germany}

\author{H.~Menke}
\affiliation{Department of Physics, University of Erlangen-N\"{u}rnberg, 91058 Erlangen, Germany}

\author{F.~Misj\'{a}k}
\affiliation{University of Ulm, Albert-Einstein-Allee 11, 89081 Ulm, Germany}

\author{D.~T.~Mantadakis}
\affiliation{Max Planck Institute for Solid State Research and Center of Integrated Quantum Network, Heisenbergstra{\ss}e 1, 70569 Stuttgart, Germany}

\author{K.~F{\"u}rsich}
\affiliation{Max Planck Institute for Solid State Research and Center of Integrated Quantum Network, Heisenbergstra{\ss}e 1, 70569 Stuttgart, Germany}

\author{E.~Schierle}
\affiliation{Helmholtz-Zentrum Berlin für Materialien und Energie, Albert-Einstein-Stra{\ss}e 15, 12489 Berlin, Germany}

\author{G.~Logvenov}
\affiliation{Max Planck Institute for Solid State Research and Center of Integrated Quantum Network, Heisenbergstra{\ss}e 1, 70569 Stuttgart, Germany}

\author{U.~Kaiser}
\affiliation{University of Ulm, Albert-Einstein-Allee 11, 89081 Ulm, Germany}

\author{B.~Keimer}
\affiliation{Max Planck Institute for Solid State Research and Center of Integrated Quantum Network, Heisenbergstra{\ss}e 1, 70569 Stuttgart, Germany}

\author{P.~Hansmann}
\email{philipp.hansmann@fau.de}
\affiliation{Department of Physics, University of Erlangen-N\"{u}rnberg, 91058 Erlangen, Germany}
\affiliation{Max Planck Institute for Chemical Physics of Solids, N{\"o}thnitzerstra{\ss}e 40, 01187 Dresden, Germany}

\author{E.~Benckiser}
\email{E.Benckiser@fkf.mpg.de}
\affiliation{Max Planck Institute for Solid State Research and Center of Integrated Quantum Network, Heisenbergstra{\ss}e 1, 70569 Stuttgart, Germany}

\date{\today}

\begin{abstract}
The recent observation of superconductivity in infinite-layer Nd$_{1-x}$Sr$_x$NiO$_2$ thin films has attracted a lot of attention, since this compound is electronically and structurally analogous to the superconducting cuprates. Due to the challenges in the phase stabilization upon chemical doping with Sr, we synthesized artificial superlattices of LaNiO$_3$ embedded in insulating LaGaO$_3$, and used layer-selective topotactic reactions to reduce the nickelate layers to LaNiO$_{2}$. Hole doping is achieved via interfacial oxygen atoms and tuned via the layer thickness. We used electrical transport measurements, transmission electron microscopy, and x-ray spectroscopy together with \textit{ab initio} calculations to track changes in the local nickel electronic configuration upon reduction and found that these changes are reversible. Our experimental and theoretical data indicate that the doped holes are trapped at the interfacial quadratic pyramidal Ni sites. Calculations for electron-doped cases predict a different behavior, with evenly distributed electrons among the layers, thus opening up interesting perspectives for interfacial doping of transition metal oxides.
\end{abstract}

\maketitle

\section{Introduction}

Ever since the discovery of unconventional superconductivity in high-$T_{\text c}$ cuprates, the search is ongoing for other $3d$ transition-metal oxides that exhibit this intriguing quantum state of matter. This is especially true for nickelates, because nickel is right next to copper in the periodic table. Recently the first observation of superconductivity in heteroepitaxially grown thin films of Sr-doped NdNiO$_2$ has raised a lot of interest.\cite{Li2019} Some time passed before these results could be reproduced experimentally, which is largely related to the delicate growth conditions.\cite{Zeng2020} Similarly to the high-$T_{\text c}$ cuprates, the phase diagram of the infinite-layer nickelates shows a dome-like doping dependence for the superconducting transition temperature $T_{\text c}$.\cite{Zeng2020, Li2020, Osada2020} The optimal doping with maximum $T_{\text c}\approx 15$\,K has been reported for Nd$_{0.8}$Sr$_{0.2}$NiO$_2$. The synthesis route of this compound requires the growth of a precursor Nd$_{0.8}$Sr$_{0.2}$NiO$_3$ thin film, which is subsequently treated by a soft-chemistry reduction with CaH$_2$. The preparation of the precursor film constitutes the first challenge, since the Sr-doped perovskite competes with Ruddlesden-Popper phases, leaving a very narrow growth window.\cite{Lee2020} The homogeneity of the Sr-dopant distribution and oxygen reduction, as well as the role of the heteroepitaxy with SrTiO$_3$ impose further challenges for the synthesis, with many details that remain yet to be explored.

\begin{figure}[b]
\center\includegraphics[width=0.8\columnwidth]{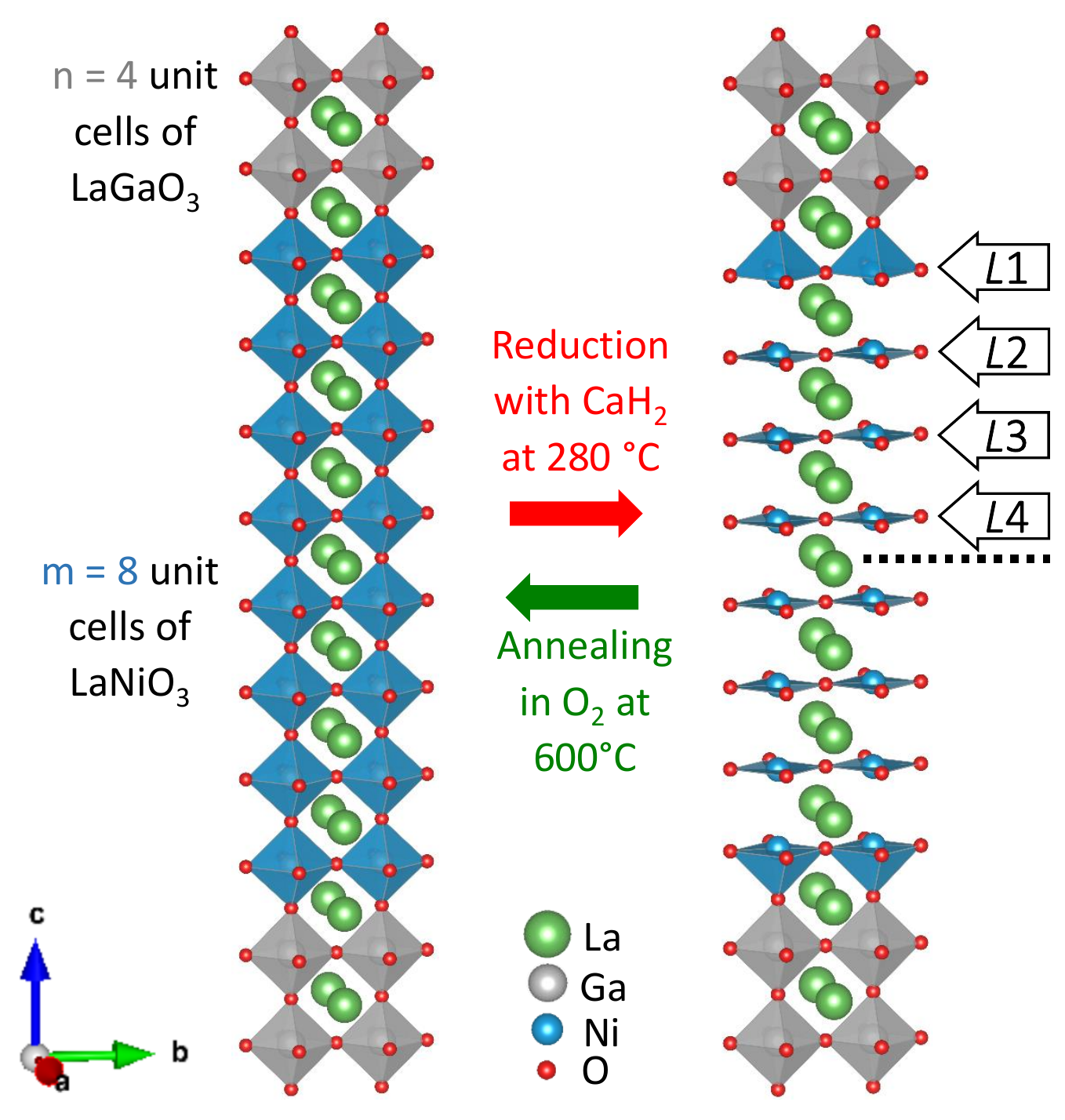}
\caption{Left: Sketch of the pristine (re-oxidized) LaNiO$_3$ (blue)- LaGaO$_3$(grey) superlattice with $m=8$, $n=4$ stacking sequences grown on a (001) SrTiO$_3$ substrate. Right: Structure stabilized by layer-selective reduction of the nickelate slab with CaH$_2$. We label individual layers within the nickelate slab by $L1$ to $L4$. The dashed horizontal line indicates the mirror plane. \label{Structure}}
\end{figure}

The soft-chemistry reduction of nickelate powders and thin films was studied for the first time several years ago.\cite{Crespin1983, Levitz1983, Hayward1999,  Crespin2005, Shimakawa2008, Kawai2009, Kaneko2009} Starting from LaNiO$_3$, the reduction to LaNiO$_2$ takes place via the formation of relatively stable phases with intermediate oxygen contents, LaNiO$_{2.75}$ and LaNiO$_{2.5}$.\cite{Moriga1995} Structurally, the reduction goes along with successive removal of apical oxygen ions, such that for the so-called brownmillerite phase LaNiO$_{2.5}$, columns of Ni ions with square-planar coordination are placed between the octahedrally coordinated nickel ions in the perovskite phase. The lattice parameters change from nearly cubic $3.84\,$\AA\ for LaNiO$_3$ to enlarged in-plane $a=3.96\,$\AA\ (perpendicular to the NiO$_4$ columns) and reduced out-of-plane $c=3.38\,$\AA\ values in LaNiO$_2$.\cite{Crespin1983, Shimakawa2008} The collapse of the $c$-axis parameter makes the reduction process trackable by x-ray diffraction (XRD) and high-resolution transmission electron microscopy (HRTEM), even for very thin films and multilayers.

Theoretical interest in this new, supposedly unconventional superconductor naturally arises from the possible similarity to, or insightful differences from high-$T_{\text c}$ cuprates.\cite{Lechermann2020,Bernardini2020,Botana2020, Zhang2020,Katukuri2020} Some theories predict much weaker magnetic correlations for the infinite-layer nickelates, and thus question their relevance for the superconducting pairing mechanism.\cite{Sawatzky2019} Other theoretical studies, however, come to different conclusions\cite{Katukuri2020} in agreement with recent results from paramagnon dispersion analysis.\cite{Lin2021} Furthermore, Ni$^{1+}$ with a formal $3d^9$ valence electron configuration is rarely found in bulk compounds, and the hybridization strength with the oxygen ligands appears to be weaker than the one of Cu$^{2+}$ in cuprates.\cite{Lee2004}

Here we report on an alternative approach to hole doping infinite-layer nickelates that takes advantage of the layer-selective topotactical reduction\cite{Matsumoto2011} of $\left(\text{LaNiO}_3\right)_m$/$\left(\text{LaGaO}_3\right)_n$ superlattices grown on $\left(001\right)$ SrTiO$_3$ substrates. The essential idea is that in a layer-selective reduction process the interfacial, apical oxygen ions remain, forming a structure depicted in Fig.~\ref{Structure}. A similar interface structure has also been suggested in recent density functional theory studies.\cite{Bernardini2020, Geisler2020} Assuming that the additional charge originating from the interfacial oxygen ions is homogeneously distributed in the LaNiO$_{2}$ layer stack, this opens the possibility of a tunable doping level, where the average nickel valence state Ni$^{1+\frac{2}{m}}$ is controlled by the number of consecutive layers $m$ (see Fig.~\ref{Structure}). A similar doping mechanism is assumed to occur in reduced Ruddlesden-Popper single crystals of (La,\,Pr)$_4$Ni$_3$O$_8$.\cite{Zhang2017} Another advantage of this approach is that the growth of the precursor superlattices is stable and the optimal conditions have been well established.\cite{Wu2013}

\section{Synthesis and characterization of superlattices}

\begin{figure}[tb]
\center\includegraphics[width=0.90\columnwidth]{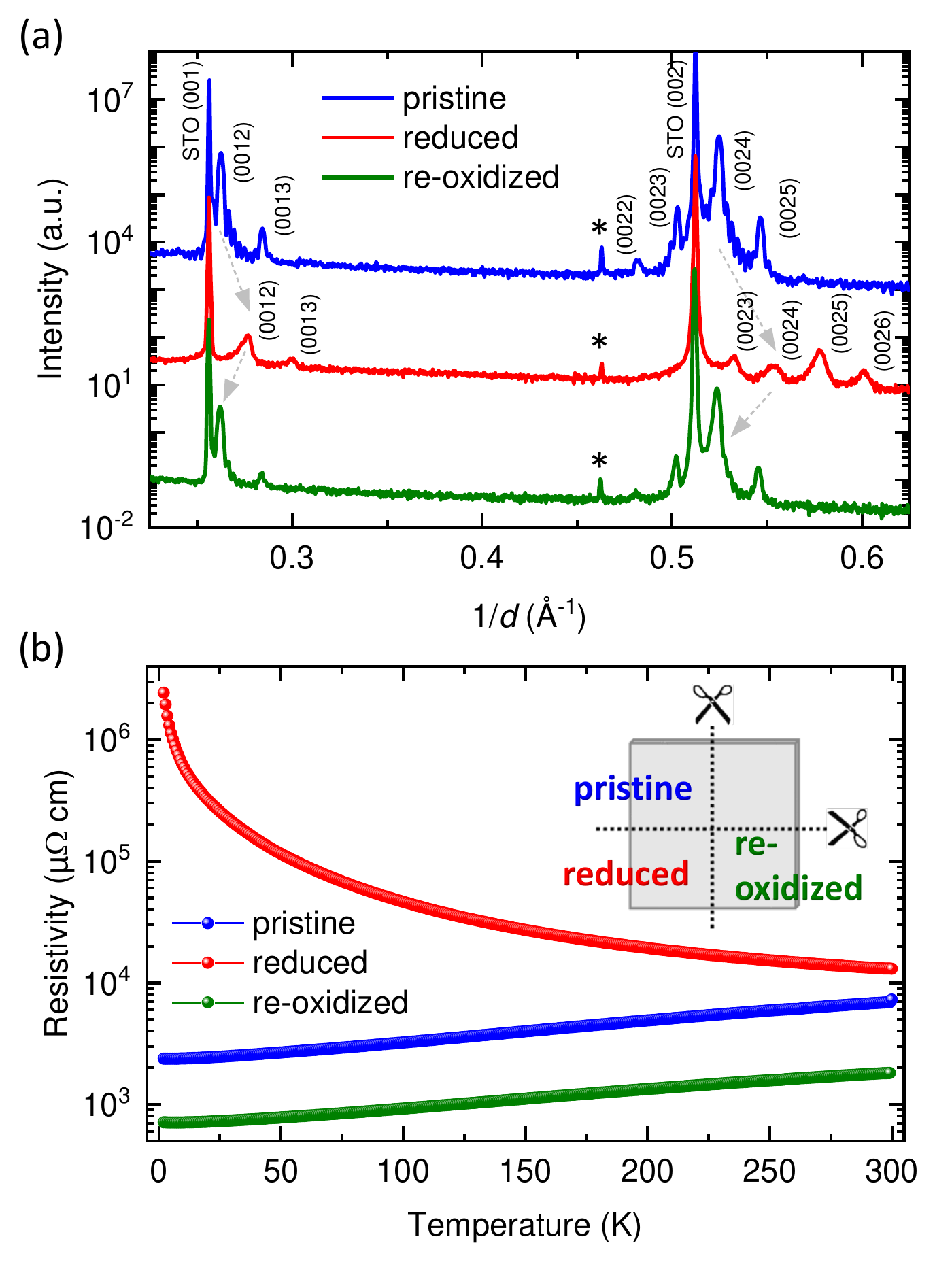}
\caption{(a) XRD patterns of a set of pristine, reduced and re-oxidized LaNiO$_{3\,(2+\delta)}$/LaGaO$_3$ superlattices including the (001) and the (002) reflections of the SrTiO$_3$ substrate. The asterisks mark peaks arising from the $K_{\beta 1}$ reflection of the main substrate peak. (b) Resistivity of the same set of samples. \label{XRD-Transport}}
\end{figure}

\begin{figure}[tb]
\center\includegraphics[width=0.87\columnwidth]{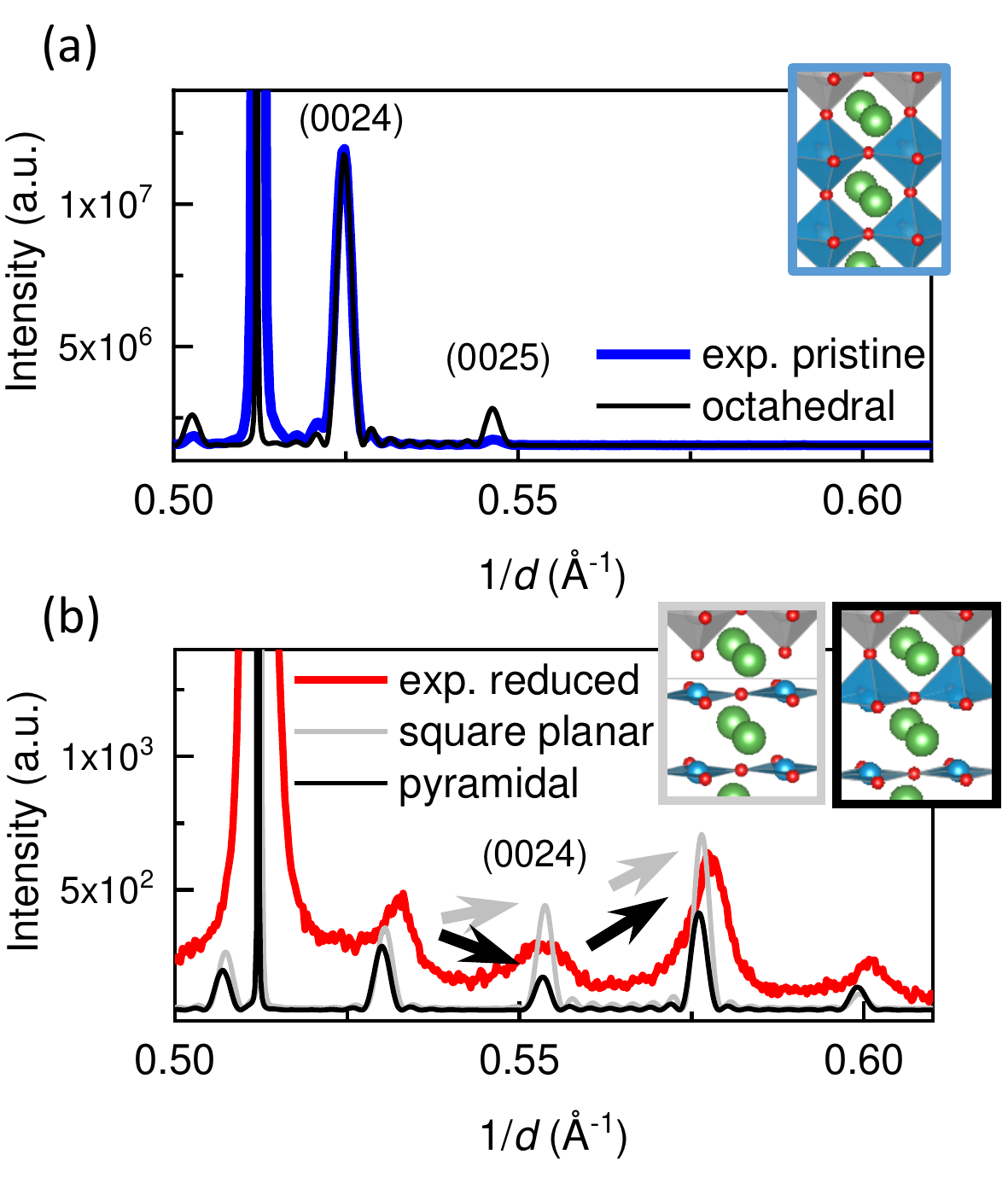}
\caption{QUAD simulations of the XRD data around the (002) SrTiO$_3$ substrate reflections for different crystal structures of the (a) pristine LaNiO$_3$ and (b) reduced LaNiO$_{2+\delta}$ superlattices. In panel (b), the simulation, of the proposed quadratic pyramidal (black box inset) interfacial nickel coordination yields a better agreement with the experimental results than the square planar (grey box inset). \label{QUAD}}
\end{figure}

In the following we focus on $\left(\text{LaNiO}_3\right)_{m=8}$/$\left(\text{LaGaO}_3\right)_{n=4}$ superlattices grown on $\left(001\right)$ SrTiO$_3$ substrates (see Fig.~\ref{Structure}). With respect to the bulk perovskite nickelate, the substrate induces a moderate tensile strain that translates into an enlarged $c$ axis parameter for the epitaxially-strained infinite-layer phase. LaGaO$_3$ was chosen to facilitate layer-selective reduction, since it is expected to be chemically more stable against oxygen removal (the Ga$^{3+}$ ions have a closed-shell configuration). Further, there is no change of the La cation sublattice across the interfaces (except from the out-of-plane La atomic distances), which removes a possible source of interfacial disorder. We emphasize that superconductivity was found in analogous infinite-layer cuprate superlattices.\cite{Castro2012}

\begin{figure}[tb]
\center\includegraphics[width=0.85\columnwidth]{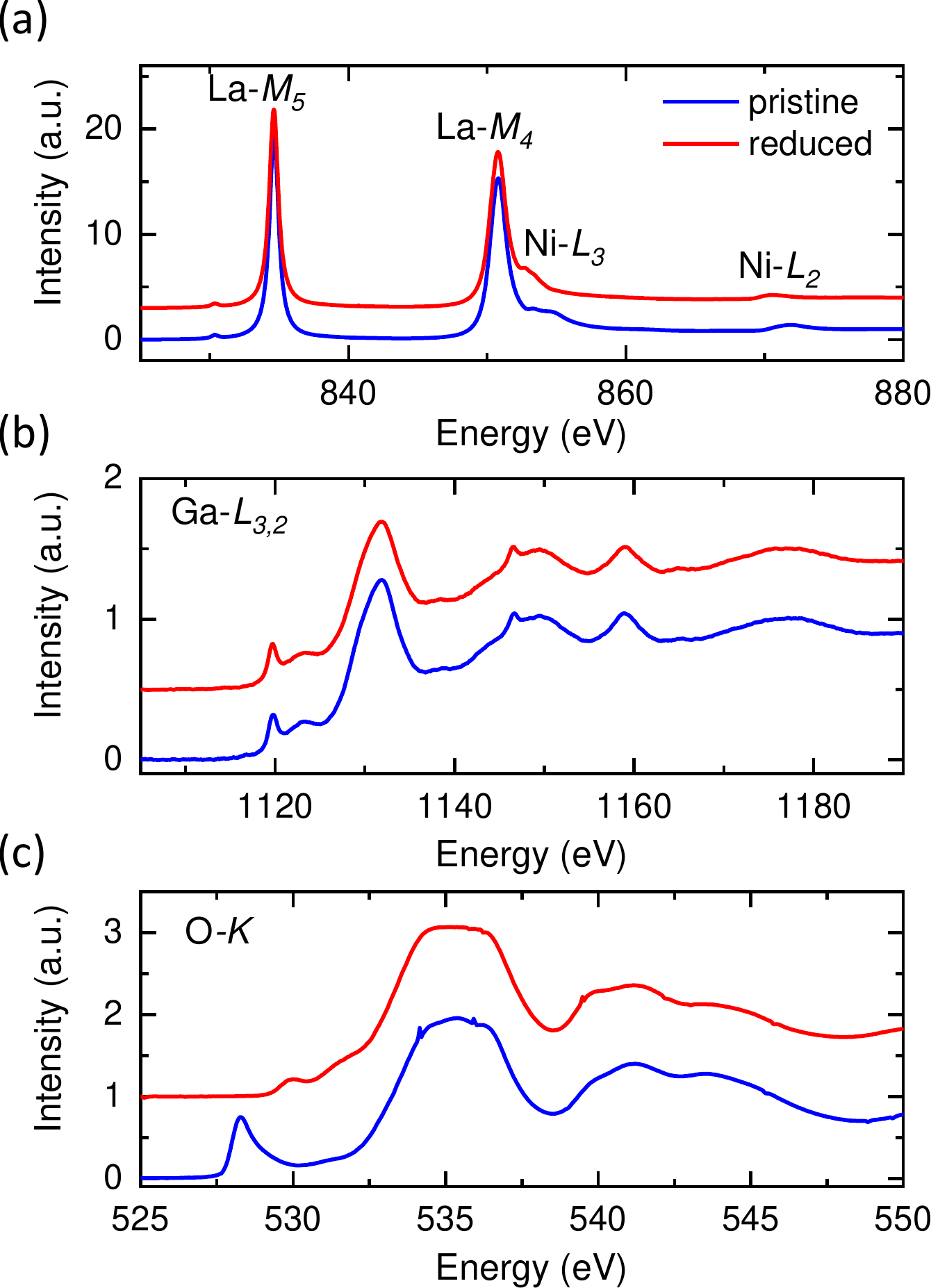}
\caption{Polarization-averaged XAS spectra ($I_{\rm average}= (2I_{E\perp c} +I_{E\parallel c})/3$; see Fig.~\ref{XASb}) of the energy range covering (a) La-$M_{5,4}$ and Ni-$L_{3,2}$ edges, (b) of the Ga-$L_{3,2}$ edges (smoothed because of the lower flux at these high energies), and (c) around the O-$K$ edge. In each panel data for pristine and reduced samples are offset for clarity. All spectra were measured in total-electron yield detection mode.\label{XASa}}
\end{figure}

\begin{figure}[tb]
\includegraphics[width=0.85\columnwidth]{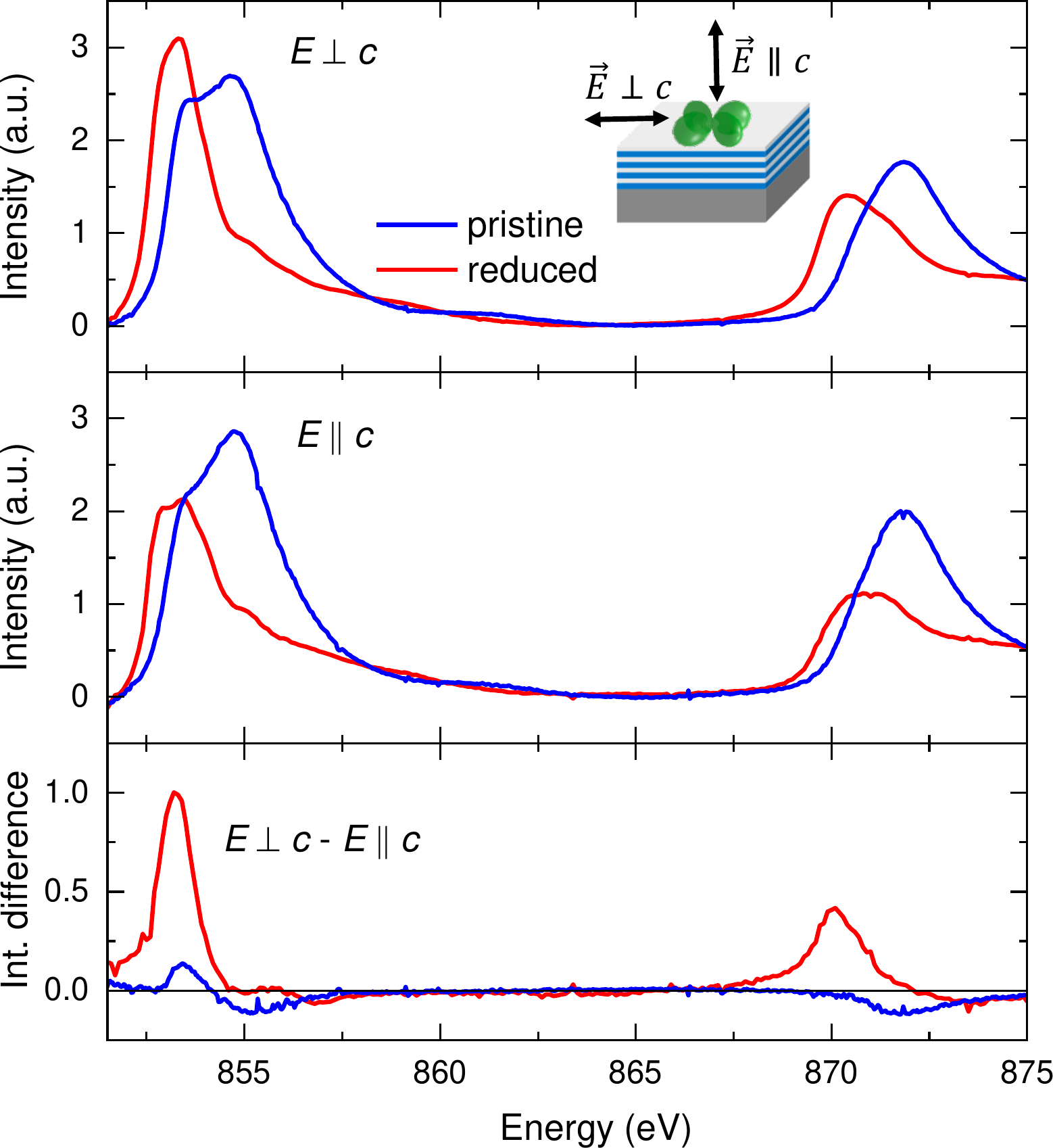}
\center\caption{Polarization-dependent XAS, measured in total-fluorescence yield for pristine and reduced samples across the Ni-$L_{3,2}$ absorption edges, where the La-$M_{4}$ lines were previously subtracted. The top and middle panel show the spectra taken with x-ray polarization $E\parallel c$ and $E\perp c$, respectively. A sketch of the geometry is shown in the inset of the top panel. The bottom panel shows the linear dichroism defined by the difference of intensities $I_{E\perp c} - I_{E\parallel c}$.\label{XASb}}
\end{figure}

The superlattices were grown by pulsed-laser deposition using high-density stoichiometric targets of LaNiO$_3$ and LaGaO$_3$ under the conditions reported in Ref.~\onlinecite{Wu2013}. In order to reduce the influence of small differences in superlattice structures, samples cut out of a single growth were either kept pristine, or reduced. Some of the reduced samples were subsequently re-oxidized (inset in Fig.~\ref{XRD-Transport}\,(b)). The as-grown superlattices were reduced using CaH$_2$ powder as a reduction agent. Specimens with dimensions $2.5\times2.5\,$mm$^2$ were placed under Ar-atmosphere inside an open aluminum foil box on top of approximately $50$ mg CaH$_2$ powder. The quartz tubes were then sealed in vacuum ($10^{-6}$~to~$10^{-7}\,$mbar) and subsequently annealed at $280^{\circ}$C for $72\,$h. X-ray diffraction and electrical transport measurements (performed in van-der-Pauw geometry) show that the reduction process is reversible (Fig.~\ref{XRD-Transport}\,(a,b)). Such reversibility of a topotactic reduction has been previously reported in LaNiO$_2$ thin films\cite{Kawai2009} and CaFeO$_2$ based superlattices.\cite{Matsumoto2011} HRTEM images show  well-ordered superlattice structures with sharp interfaces (see Appendix~\ref{TEM}) and confirm the layer selectivity of the reduction process. The collapse of the $c$-axis parameter only occurs in the LaNiO$_{2+\delta}$ layers. Similar lattice parameter contraction of reduced Nd$_{1-x}$Sr$_x$NiO$_2$ thin ?lms was observed by annular bright-field scanning-TEM, close to the substrate.\cite{Lee2020} The averaged bilayer thicknesses of $46.2(2)$ and $44.0(6)$\,\AA~for pristine and reduced samples respectively are in good agreement with the corresponding XRD values of 45.7(1) and $43.4(1)$\,\AA.

While the electrical transport of pristine and re-oxidized samples exhibits metallic behavior closely analogous to bulk LaNiO$_3$, the resistivity of the reduced superlattices shows a semiconducting temperature dependence (Fig.~\ref{XRD-Transport}\,(b)). We found that the conductivity of the re-oxidized samples is reproducibly higher than that of the pristine ones. We attribute this finding to the presence of oxygen vacancies in the as-grown samples. Although samples were annealed in 1-bar oxygen atmosphere at $690^{\circ}$C as the last step of the PLD growth procedure, this annealing of the pristine samples is less efficient than reduction and subsequent annealing. While the filling of individual, isolated oxygen vacancies in the annealing process of pristine samples is limited by oxygen diffusion, this restriction is lifted when the apical oxygen sites are first emptied and then refilled.

Detailed analysis of the (00$L$) x-ray diffraction scans and comparison with patterns calculated by the simulation software QUAD\cite{Macke2016} (Fig.~\ref{QUAD}\,(a,b)) confirmed the interfacial structure depicted in Fig.~\ref{Structure}. In particular, the relative intensities of the $L=23, 24, 25$ superlattice reflections (high-low-high) are in good agreement with the simulation of quadratic pyramidal Ni-O coordination, while the monotonically increasing intensity pattern predicted for square-planar coordination disagrees with the experimental data. This characteristic intensity pattern was observed in various samples after topotactic reduction.

\section{X-ray Absorption Spectroscopy \label{XASs}}

To study modifications of the electronic configuration of the different ions upon reduction we examined soft x-ray absorption spectra (XAS) across several relevant absorption edges at room temperature. The measurements were carried out at the UE46 PGM-1 beamline of the BESSY-II synchrotron at Helmholtz-Zentrum Berlin. The spectra around the La-$M_{5,4}$ ($3d\rightarrow 4f$) and Ni-$L_{3,2}$ ($2p\rightarrow 3d$) absorption edges are shown in Fig.~\ref{XASa}\,(a). Compared to the pristine piece of our sample, the center of mass of the Ni-$L$ white lines of the reduced sample is shifted to lower energies by about 1.5\,eV, which indicates a reduced Ni valence state. On the contrary, the La-$M$ and Ga-$L$ lines remain unchanged within in our energy resolution. The absence of a shift and nearly identical line shapes of the Ga-$L$ edge spectra before and after reduction are strong indications of the layer selectivity of the reduction (Fig.~\ref{XASa}\,(b)). The comparison of spectra at the O-$K$ edge, on the other hand, indicates a strongly modified Ni-O hybridization (Fig.~\ref{XASa}\,(c)). Perovskite LaNiO$_3$ is a negative charge-transfer material, where the largest contribution to the Ni ground state configuration is $3d^8\underline{L}$ (where $\underline{L}$ denotes a ligand hole).\cite{Green2016} This strong hybridization with nearly one hole in the oxygen $2p$ states results in a pronounced O-$K$ edge pre-peak around 528\,eV. In the spectrum of the reduced superlattice, the spectral weight shifts towards higher energies and decreases considerably, indicating a loss of hybridization. A related effect has been observed in resonant inelastic x-ray scattering experiments on NdNiO$_2$ and in model calculations predicting that the infinite-layer nickelates fall into the Mott-Hubbard regime, where the lowest-energy electron-addition states are of Ni $3d$ character.\cite{Hepting2020,Jiang2020} Such a two-dimensional, narrow-band electron system is prone to localization, in agreement with the electronic transport data of Fig.~\ref{XRD-Transport}(b).

Next we turn to the polarization dependence of the XAS spectra. The spectra of pristine LaNiO$_3$ are nearly isotropic, because the two $d$-shell holes on the octahedrally coordinated Ni$^{3+}$ ions are approximately evenly distributed over the $d_{x^2 - y^2}$- and $d_{3z^2 - \mathbf{r}^2}$-based orbitals.\cite{Wu2013} In reduced samples, on the other hand, a strong difference between XAS spectra with x-rays polarized along (I$_{E\parallel c}$) and perpendicular to the $c$-axis (I$_{E\perp c}$) is expected to arise from the single hole on the Ni$^{1+}$ ions in planar coordination, which favors the $d_{x^2 - y^2}$ orbital (inset in Fig.~\ref{XASb}). In line with this expectation, we only observed a weak negative dichroism (defined as $I_{E\perp c}-I_{E\parallel c}$) in the pristine samples (blue lines in Fig.~\ref{XASb}), which can be attributed to combined effects of epitaxial tensile strain from the SrTiO$_3$ substrate and electronic confinement by the insulating LaGaO$_3$ spacer layers.\cite{Wu2013} As expected from the change in electron filling and crystal field, the dichroism of the reduced sample is greatly enhanced and shows a change in sign. Its line shape is remarkably similar to the dichroism of Cu$^{2+}$ in high-$T_{\text c}$ cuprates, albeit reduced in size.\cite{Chen1992} This quantitative difference is possibly related to the unevenly distributed charge across the LaNiO$_{2+\delta}$ layer stack. As we will explain in the following section, the \textit{ab initio} calculations indicate that the self-doped holes at the interface are trapped and contribute with a strongly reduced linear dichroism to the layer-averaged spectra.

\section{Theory: Model and Methods}

\begin{figure*}[tb]
\center\includegraphics[width=\textwidth]{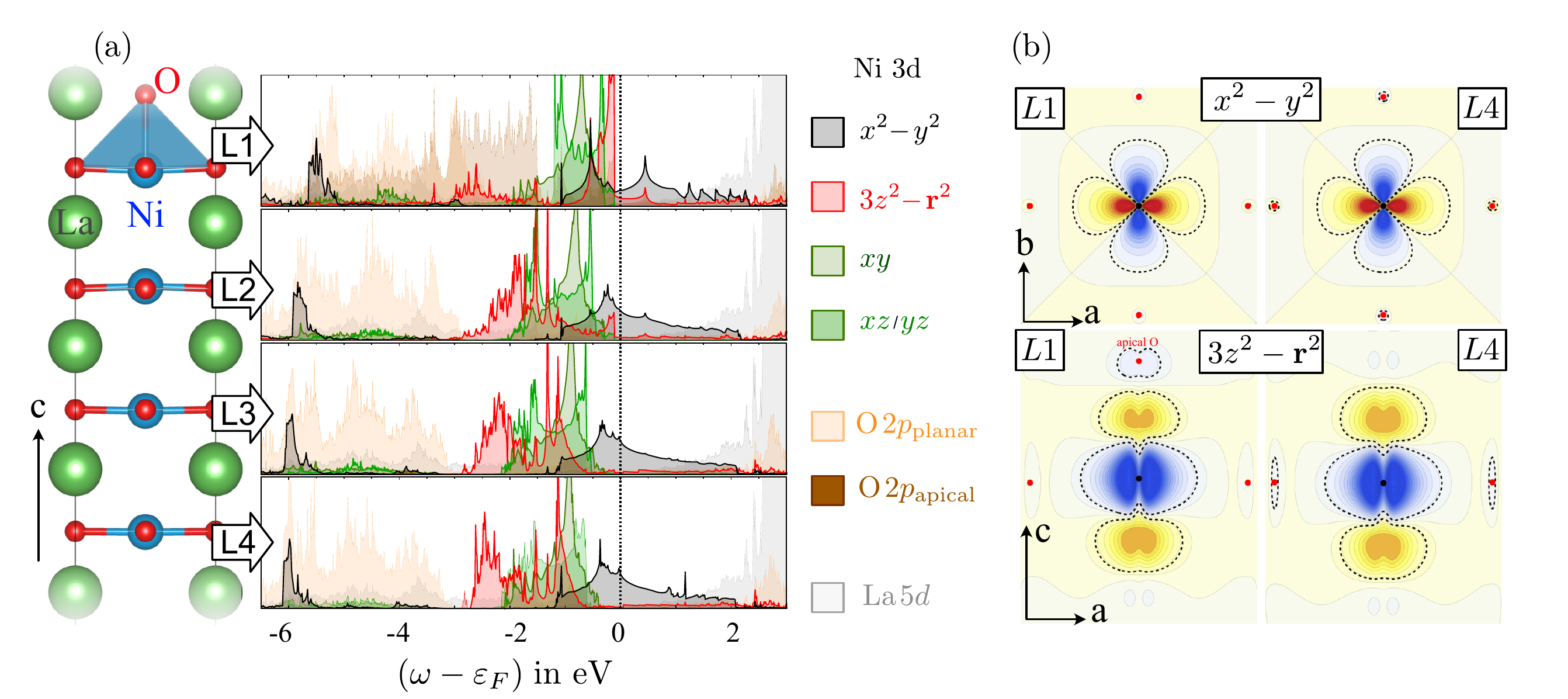}
\caption{a) Partial density of states of the inequivalent LaNiO$_{2+\delta}$ layers in the 8-layer stack with respect to electronic orbitals near the Fermi level (except for the La-$5d$ pDOS, which
is averaged over all La sites) from DFT calculations. b) Contour plot of Wannier functions with planar $x^2-y^2$ symmetry in the $ab$-plane (upper panels), and axial $3z^2-\mathbf{r}^2$ symmetry in the $ac$-plane (lower panels) for interface- ($L1$) and innermost layer ($L4$). Black (red) dots indicate the position of Ni (O) ions. The dashed line marks the contour value of $0.05$. \label{fig:DFT}}
\end{figure*}

To gain further insight into the layer- and orbital-resolved correlated electronic structure we performed \emph{ab initio} density functional theory (DFT) plus dynamical mean-field theory (DMFT) calculations. We relaxed the ionic positions of the reduced unit-cell shown on the right-hand side in Fig.~\ref{Structure} with the Vienna Ab initio Simulation Package (VASP) code\cite{Kresse1993} using the generalized gradient approximation (GGA) functional.\cite{Perdew1996} Subsequent DFT self-consistency was performed using the full-potential local-orbital (FPLO) code.\cite{Koepernik1999,Opahle1999} For the Brillouin zone (BZ) integration we used the tetrahedron method with a $12 \times 12 \times 1$ $\mathbf{k}$-mesh. For the DMFT calculations we downfolded the converged Kohn-Sham eigenstates to a basis of $195$ orbitals (per spin) including (i) the $4s$- and $3d$-shells of all Ni ions, (ii) the $2p$-shell of all O ions, and (iii) the $5d$-shell of all La ions. We note that especially the explicit treatment of Ni $4s$ as well as La $5d$ degrees of freedom significantly improved the localization of the effective $3d$ Wannier functions. In Fig.~\ref{fig:DFT}(b) we show examples of localized $d_{x^2-y^2}$ and $d_{3z^2-\mathbf{r}^2}$ Wannier orbitals at the interface $L1$ and the inner layer $L4$.

The dynamical mean-field calculations were performed by converging four (one for each nonequivalent Ni site) auxiliary impurity problems in full charge self-consistency. Each one of the impurity problems contains the full $3d$ shell of the respective Ni ion and was solved for a fully $\mathrm{SU}(2)$ symmetric Kanamori operator with Hubbard $U=8.0$~eV, Hund's coupling $J_H=0.8$~eV, and the assumption of fixed inter-orbital interaction $U'=U-2J_H$,\cite{Oles1983, Georges2013} which is a common approximation for the full local Coulomb operator (See Appendix~\ref{app:DMFT_Kanamori} for the full Hamiltonian). The value of $U$ might appear unusually large, but this is merely a consequence of the large basis containing O-$2p$, La-$5d$, and Ni-$4s$ orbitals, which accounts for screening effects of the Coulomb interaction explicitly. Generally speaking, the more orbitals are included in the calculation, the more the value of $U$ approaches the bare atomic Coulomb repulsion. For our calculations we used the w2dynamics code\cite{Wallerberger2019} and its associated continuous-time quantum Monte Carlo solver in the hybridization expansion version. To account for the appearance of the Hartree term in both DFT and DMFT we employ the ``fully-localized limit'' double-counting correction.\cite{Held2007} All calculations were performed at a temperature of $T = 1/50 (\mathrm{eV}^{-1})\approx 232$\,K. Following common procedure the total charge of the lattice model remains fixed within the DMFT self-consistent calculation, which allows for a well-defined simulation of doping in the correlated system. All calculations were performed in the paramagnetic phase, i.e.\ magnetic order is suppressed to overcome the ordering tendency inherent to a spatial mean-field method like DMFT. The analytical continuation to calculate the DMFT spectral function shown in Fig.~\ref{fig:DMFT_spectra} was performed by using the maximum entropy method (MaxEnt) as implemented in Ref.~\onlinecite{MaxEntCode}. Further calculation details can be found in Ref.~\onlinecite{Mantadakis2019}.

\section{Non-interacting reference model}
We start our discussion on the level of the non-interacting reference model derived from density functional theory. In Fig.~\ref{fig:DFT}(a) we show the layer- and orbitally resolved partial density of states (pDOS) around the Fermi level ($\varepsilon_F$). We find for all layers the Ni $3d$ states to be partially filled and in-between the O-$2p$ and the La-$5d$ states. Another common observation for all layers is that the Ni-$d_{x^2-y^2}$ weight is the most prominent at and above $\varepsilon_F$. Even though we find some of the $d_{x^2-y^2}$ weight at around $-6\,\mathrm{eV}$ in a covalent bond with oxygen, its overall hybridization with the O-$2p$ states is rather weak (e.g.\ in comparison to charge-transfer systems like high-$T_c$ cuprates). This is in accordance with the experimental O-$K$ edge spectra discussed in section~\ref{XASs}.

A closer look at the pDOS in Fig.~\ref{fig:DFT}(a) below $\varepsilon_F$ reveals significant differences between the electronic structure of interface $L1$ and the inner layers $L2$--$L4$. This is due to ligand-field effects involving the apical oxygen at the interface and the apical oxygen vacancies in the inner layers. Due to its extension along the $c$-axis, it is not surprising that the $d_{3z^2-\mathbf{r}^2}$ orbitals are affected the most. The strong $\sigma$-hybridization with the apical oxygen states in the interface layer $L1$ leads to a large peak between $-1.0$ to $-0.5\,\mathrm{eV}$ corresponding to the antibonding combination of $d_{3z^2-\mathbf{r}^2}$ and apical oxygen $2p_z$. As a secondary effect, driven by non-local hybridization of the axial state with the planar $d_{x^2-y^2}$ orbitals, we also find an additional peak in the $d_{x^2-y^2}$ pDOS in the interface layer $L1$ at about $-1\,\mathrm{eV}$, which is absent in the inner layers. Interfacial $L1\colon d_{xz} / d_{yz}$ orbitals are also affected by the apical oxygen (albeit on a smaller energy scale) due to $\pi$-hybridization with apical $p_x$- and $p_y$-orbitals which brings them closer to $\varepsilon_F$ for the interface layer.

The apical oxygen vacancies in the inner layers lead to a different ligand-field mechanism for the $d_{3z^2-\mathbf{r}^2}$ states. We observe that the oxygen vacancy is not simply an ``absence of oxygen'' but provokes the extension of especially the La-$5d$ orbitals into the vacancy site. Their probability density at the vacancy site is large enough that a single-particle basis transformation to a ``bond-like'' state which has been referred to as ``zeronium'' orbital\cite{Rompa1984} leads to an insightful perspective: when apical oxygen orbitals (which are full in the ionic limit) are replaced by zeronium orbitals (which are empty in the ionic limit), the $d_{3z^2-\mathbf{r}^2}$ states around $\varepsilon_F$ no longer form antibonding states with apical oxygen $2p_z$, but instead they form \emph{bonding} zeronium-$d_{3z^2-\mathbf{r}^2}$ states, which is reflected in a dramatic lowering of the corresponding $d_{3z^2-\mathbf{r}^2}$ pDOS in the inner layers and a small, but non-negligible increase in the weight of $d_{3z^2-\mathbf{r}^2}$ to energies well above $\varepsilon_F$. Interestingly, the electronic structure of the interface Ni site (in $L1$) shows features originating from both ligand-field effects, the one with apical oxygen and the one with zeronium ligand.

The layer-dependent ligand field effects are also reflected in the orbital occupations projected into the Ni-$3d$ subspace listed in Tab.~\ref{tab:occupations_DFT} of  Appendix~\ref{densities}. The small difference in total $3d$ occupations of central and interface layers of only $n_{\mathrm{tot}}(L4)-n_{\mathrm{tot}}(L1) = 0.06$ might seem surprising in light of the strong layer dependence of the pDOS. The reason for this small difference is a ``damping'' of the density differences arising from the aforementioned ligand field effects: Expanding the electronic configurations around the ionic limit with a layer-dependent ansatz looks like this:
\begin{multline}
  \lvert\Psi_{L1}\rangle=\\
   \alpha_1 (\mathrm{Ni}\,3d^8)
  + \beta_1 (\mathrm{Ni}\,3d^9 \underline{L})
  + \gamma_1 (\mathrm{Ni}\,3d^7 z)
  + \ldots
\end{multline}
for the interface and
\begin{multline}
  \lvert\Psi_{L2\text{--}L4}\rangle=\\
  \alpha_{2\text{--}4} (\mathrm{Ni}\,3d^9)
  + \beta_{2\text{--}4} (\mathrm{Ni}\,3d^{10} \underline{L})
  + \gamma_{2\text{--}4} (\mathrm{Ni}\,3d^8 z)
  + \ldots
\end{multline}
for the inner layers. Here $\underline{L}$ denotes an (oxygen) ligand-hole and $z$ a (zeronium) ligand electron. The hybridization with apical oxygen at the interface layer increases the Ni-projected charge due to an enhancement of $\beta$ (i.e.\ the formation of oxygen holes), while hybridization with zeronium in the inner layers enhances $\gamma$ and thus depletes it. The density analysis shows the limited value of projected density operators and underlines the need to use a configuration interaction language rather than ``occupations''.

In summary we find strong and qualitative differences of the electronic structure for orbitals with extension along the c-axis. The most relevant mechanism is the (covalent) ligand-field effects for the $d_{3z^2-\mathbf{r}^2}$ orbital whose spectral weight is pushed towards the Fermi level in $L1$ due to the apical oxygen and pushed away from the Fermi level in $L2$--$L4$ due to its overlap with zeronium states. We will now see, that these layer-dependent orbital polarization effects on the single-particle level are even further enhanced when electronic interactions are considered beyond the static mean-field level.

\section{Dynamical Mean-Field Theory}
\begin{figure*}[tb]
  \centering
  \includegraphics[width=\textwidth]{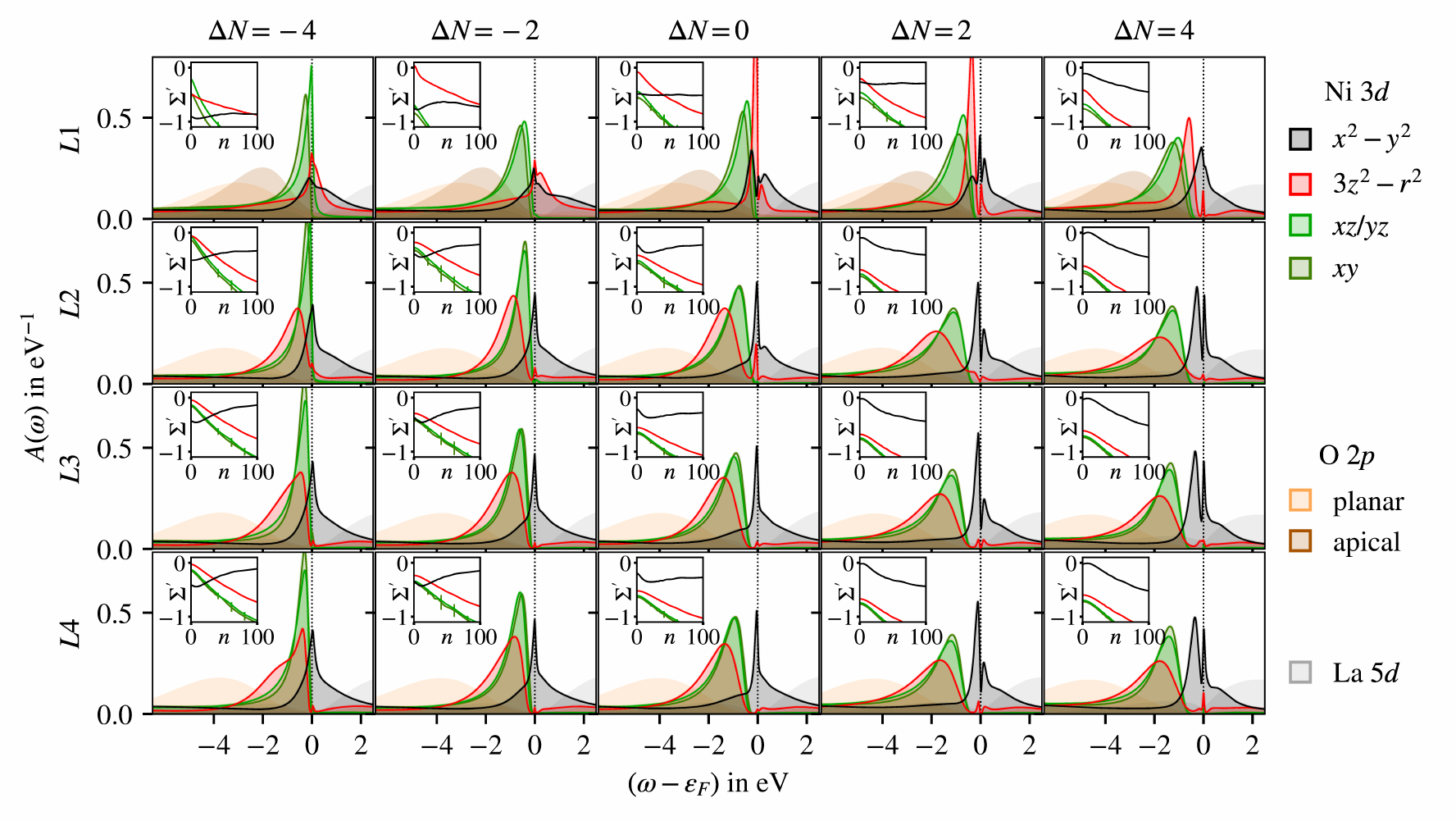}
  \caption{Layer-($L1-L4$) and orbitally resolved $\mathbf{k}$-integrated spectral function A$(\omega)$ from DMFT for interaction values $U = 8\,\mathrm{eV}$ and $J = 0.8\,\mathrm{eV}$ at $T = 1/50 (\mathrm{eV}^{-1})\approx 232$\,K. The columns correspond to different total electron fillings, where negative (positive) $\Delta N$ corresponds to hole (electron) doping. The insets show the orbital dependent real part of the DMFT self-energy $\Sigma'(i\omega_n)$ on the Matsubara frequency axis ($\omega_n = (2n+1)\pi k_B T$, $n \in \mathbb{Z}$). The energy range and the color code for the plots are identical to the ones used in Fig.~\ref{fig:DFT}. \label{fig:DMFT_spectra}}
\end{figure*}

We start the discussion of our DMFT calculations (for $U = 8\,\mathrm{eV}$ and $J = 0.8\,\mathrm{eV}$ at $T = 1/50 (\mathrm{eV}^{-1})\approx 232$\,K) with the layer- and orbital resolved single-particle spectral functions, which are shown in Fig.~\ref{fig:DMFT_spectra} for different values of total filling ($\Delta N = 0$ corresponds to the reduced heterostructure as displayed in Fig.~\ref{Structure} without any extra charge).

\subsection{Nominal doping}
Comparing the DMFT spectra of the system at nominal doping ($\Delta N = 0$) to the partial DOS from DFT in Fig.~\ref{fig:DFT} reveals two major effects originating from local correlations: (i) a re-normalization of the quasi-particle (QP) masses and (ii) an enhancement of the layer dependent orbital polarizations. Effect (i) is the expected sharpening of the spectra at the Fermi level. The corresponding (orbital dependent) mass ($m$) re-normalization $Z_\alpha\equiv m^\text{QP}_\alpha/m^\text{free}_\alpha$ depends on layer and orbital (index $\alpha$) and can be estimated from the self-energy (for details see App.~\ref{app:DMFT_sigma}). At nominal doping the inner $d_{x^2-y^2}$ QPs have $4$~to~${5 \times m^\text{free}}$, while at the interface we find ${2.5 \times m^\text{free}}$. A closer look at the interface reveals also a second QP peak of $d_{3z^2-\mathbf{r}^2}$-character with $Z_{3z^2-\mathbf{r}^2}\approx 0.5$. This is a consequence of (ii), i.e. the enhanced layer-dependent orbital polarization of the Ni $3d$ shell.  In the interface layer $L1$ the spectral weight of the $d_{3z^2-\mathbf{r}^2}$ orbital is pushed across the Fermi level and is the dominant contribution to the sharp QP peak around $\varepsilon_F$.  In the inner layers, in contrast (especially $L3$ and $L4$), the spectral weight of the $d_{3z^2-\mathbf{r}^2}$ orbital remains well below $\varepsilon_F$ and the QP peak has almost pure $d_{x^2-y^2}$ character.

The results for the spectral weight transfer from the MaxEnt method are supported by data obtained directly on the Matsubara axis.  The stochastic nature of the MaxEnt method always incurs an uncertainty for the analytical continuation, which is not present for quantities on the Matsuabra axis, which are produced directly by the quantum Monte Carlo solver.  In the insets of Fig.~\ref{fig:DMFT_spectra} we show the real part of the DMFT self-energy $\Sigma'(i\omega_n)$ (w.r.t.\ the chemical potential $\mu=0$) for the correlated $d$-orbitals with error bars plotted on the Matsubara frequency axis (color code as for the spectral functions). The extrapolation of $\Sigma'(i\omega_n\rightarrow 0)$ provides an estimate of the orbital-dependent spectral weight shifts at small energies. This self-energy driven spectral polarization changes sign from the interface (where the extrapolation of self energy for the $d_{3z^2-\mathbf{r}^2}$ orbital is $\approx 0.5\,\mathrm{eV}$ larger than for the other orbitals) to the inner layers, where the $d_{x^2-y^2}$ weight is shifted up in energy instead. This means that local correlation effects \emph{enhance} the tendencies of the single particle DFT calculations. The origin of this enhancement is the Hund's coupling and has been studied in detail for generic $dp$-models in DMFT:\cite{Parragh2013} if orbitals are close in energy, additional Hund's energy can be gained by equalizing orbital occupations and maximizing total spin, which is what we observe at the interface.\footnote{We stress the many-body nature of this effect, which is reflected in the dynamic nature of $\Sigma$ and the qualitative difference of orbital shifts at small (around $\varepsilon_F$) and large frequencies (corresponding to a static mean-field Hartree-limit of the self-energy).} For the inner layers, however, high-spin configurations are too high in energy to begin with and Hund's coupling is less important. For completeness, the projected densities at the Ni site from DMFT calculations are provided in Tab.~\ref{tab:occupations_DMFT} of Appendix~\ref{densities}.

\subsection{Hole and electron doping}

\begin{figure*}[tb]
  \centering
  \includegraphics{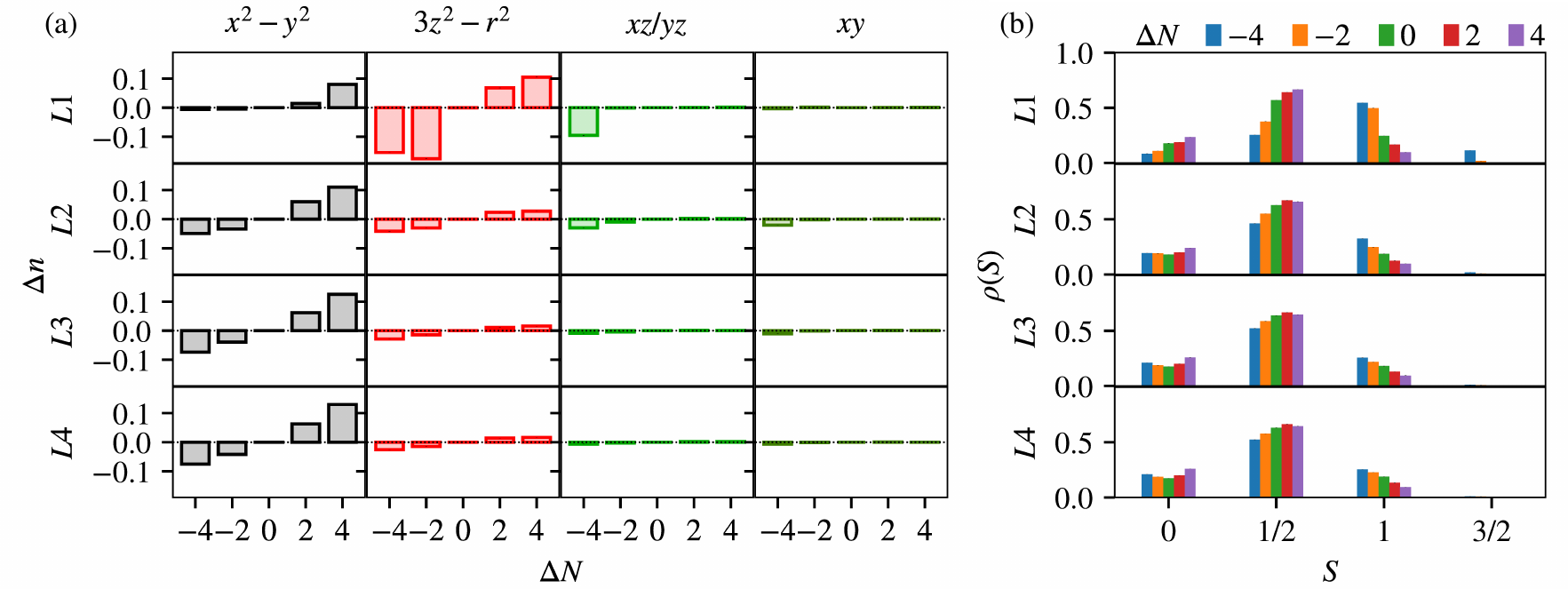}
\caption{(a)~Layer- and orbital-resolved distribution of the density~$\Delta n$ at the inequivalent Ni sites.  (b)~Layer-resolved spin density~$\rho(S)$ as determined by projecting the many-particle density matrix into the basis of the $S^2$ operator.\label{fig:DMFT_histograms}}
\end{figure*}

To improve our understanding, we consider the effect of extra charge in the heterostructure. To this end we consider the addition of two and four holes or electrons per unit cell, i.e.\ a change in the total number of electrons of $\Delta N \in \{ -4, -2, 2, 4 \}$ in comparison to the nominal doping $\Delta N=0$. The corresponding spectra are shown in Fig.~\ref{fig:DMFT_spectra}. For the inner layers $L2$--$L4$ we find singe-band behavior: doping the $d_{x^2-y^2}$ band away from half-filling decreases the mass-renormalization of the QPs (see $Z$-factors in Fig.~\ref{fig:DMFT_self_energies} of Appendix \ref{selfenergies}). The same is true also for the interface layer $L1$ in the electron-doped case.

The spectra of the interface-layer on the hole-doped side are, however in stark contrast to this correlated single-band picture: For $\Delta N =-2$ we observe in the interface layer an increased transfer of spectral weight of the $d_{3z^2-\mathbf{r}^2}$ orbital and an even larger splitting in the real-part of self-energy to the $d_{x^2-y^2}$ orbital compared to the nominal doping $\Delta N=0$ case. Due to this shift, a large amount of the extra holes is trapped at the interface. This is also true for even larger hole-doping $\Delta N=-4$, where instead of an even distribution across the layers the holes migrate to interface $d_{xz/yz}$ orbitals.

To understand the mechanism leading to the trapping of holes in the interface layer, we provide histograms of the layer- and orbital-resolved projected densities at each Ni site, relative to the value at nominal doping in Fig.~\ref{fig:DMFT_histograms}(a) and of the configurations of the Ni local spin moments in Fig.~\ref{fig:DMFT_histograms}(b). On the electron-doped side the density histograms in Fig.~\ref{fig:DMFT_histograms}(a) confirm that the additional electrons distribute evenly across layers. On the hole-doped side, in contrast, the extra carriers are trapped at the interface layers initially in the $d_{3z^2-\mathbf{r}^2}$ orbital and then in the $d_{xz/yz}$ orbital, once the $d_{3z^2-\mathbf{r}^2}$ orbital has become half-filled. The reason for this trapping can be understood by considering the spin configurations of the local Ni moment in Fig.~\ref{fig:DMFT_histograms}(b). For $\Delta N=-2$ the center of the configuration distribution shifts from $S = 1/2$ to $S = 1$, which is representative for the formation of the high-spin state in the $d_{3z^2-\mathbf{r}^2}$ orbital to gain Hund's coupling energy. At $\Delta N=-4$ the maximum at $S = 1$ becomes more pronounced and we observe a finite weight of $S = 3/2$ configurations (consistent with added holes of $d_{xz/yz}$ character). The true many-body nature of this local moment formation is also reflected in the mass of interface $L1$ QPs: instead of a decreased mass re-normalization (like we observe for the inner layers) the Hund's correlation dominated $L1$ QPs show $Z$-factors as low as $Z_{3z^2-\mathbf{r}^2}\approx 0.06$ (i.e.\ a~QP mass enhancement by a factor $\approx 16.6$) as estimated from Fig.~\ref{fig:DMFT_self_energies}. This observation indicates the creation of a strongly correlated Hund's metal\cite{deMedici2011} in the $L1$ interface, which traps most of the doped holes. Finally we note that the distribution of the extra charge into O-$2p$ and La-$5d$ states is negligible.

In summary our theoretical analysis indicates that the supposedly homogenously self-doped heterostructure has a strongly layer-dependent electronic structure. Ligand-field effects due to apical oxygen and apical vacancies on the single-particle level are enhanced by local correlation effects. Careful analysis of spectral functions, projected densities, many-body configuration histograms, and the dynamic self-energies indicate the formation of a strongly correlated Hund's metal at the interface layer. In the concomitant formation of local high-spin configurations, doped holes are trapped at the interface, rather than being distributed evenly across the heterostructure. In contrast, doped electrons are distributed evenly across the layers, filling up $d_{x^2-y^2}$ states.

\section{Summary and Conclusions}
In summary, we have shown that reversible, layer-selective removal of oxygen ions yields LaNiO$_{2+\delta}$-LaGaO$_3$ superlattices with well-ordered infinite-layer nickelate stacks, and we used x-ray diffraction and transmission electron microscopy to characterize the structure. Based on our combined results from electrical transport, x-ray spectroscopy, and \textit{ab initio} calculations, we draw the following conclusions: (i) The oxygen reduction of the superlattice is accompanied by a change in the electronic structure, which is characterized by a transition from a negative charge-transfer to a Mott-Hubbard type system. The shift of the O-$p$ derived bands to energies well below the occupied Ni-$d$ states in DFT+DMFT is supported by our observations from O-$K$ edge XAS, where the $d^8\underline{L}\rightarrow d^{8}$ pre-peak of the pristine samples is completely suppressed and the only remaining $d^9\underline{L}\rightarrow d^{10}$ transitions in reduced samples are weak and hidden under the continuum part of the spectrum. Experiments on undoped infinite-layer LaNiO$_2$ thin films came to a similar conclusion.\cite{Hepting2020} (ii) The sharp peaks at $\varepsilon_F$ in the DMFT spectra indicate the presence of heavy quasi-particles. The small amount of coherent spectral weight and sizable mass-enhancement factors indicate that the system is very close to a insulating state which is consistent with the experimental findings of a semiconducting electronic transport behavior. Either additional electronic correlations that were not taken into account in our DFT + DMFT calculations can lead to a Mott-isolating state, or localization is caused by residual disorder in the experimentally realized superlattices. A combination of both effects is equally possible. (iii) The layer-averaged, overall dominant Ni-$d_{x^2-y^2}$ orbital character of the electron-addition states calculated in DMFT qualitatively agrees with the spectral shape and sign of the observed linear dichroism measured by  Ni-$L$ edge XAS. For the inner layers $L2$--$L3$ the DMFT results are reminiscent of the single-band Hubbard model that possibly hosts superconductivity upon doping. (iv) Despite having a very similar self-doping level as the superconducting Nd$_{0.8}$Sr$_{0.2}$NiO$_2$ films reported in Ref.~\onlinecite{Li2019}, the superlattices do not exhibit superconductivity down to $2$\,K (Fig.~\ref{XRD-Transport}). The electronic-structure calculations yield an explanation for this striking difference in terms of Hund's coupling stabilized trapping of the self-doped holes at the superlattice interfaces. Based on this understanding, the DMFT results provide a promising outlook for this superlattice structures upon electron-doping, where a uniform distribution of the doped electrons across the nickelate layer stack should be possible. More generally, our combined experimental and theoretical results provide an interesting perspective for the interfacial doping approach of transition-metal oxide superlattices.

\begin{acknowledgements}
We acknowledge financial support by the Center for Integrated Quantum Science and Technology (IQ$^{\rm ST}$) and the Deutsche Forschungsgemeinschaft (DFG, German Research Foundation): Projektnummer 107745057~-~TRR 80 and Projektnummer 323667265. This work is supported by a project funded by the Carl Zeiss foundation. We thank HZB for the allocation of synchrotron radiation beamtime. We thank G.~A.~Sawatzky, K.~Foyevtsova, R.~Merkle, and M.~Klett for inspiring and very insightful discussions. We thank Manuel Mundszinger for FIB sample preparation.
\end{acknowledgements}


\newpage
\appendix

\section{HRTEM Measurements \label{TEM}}
\begin{figure}[b]
\center\includegraphics[width=0.99\columnwidth]{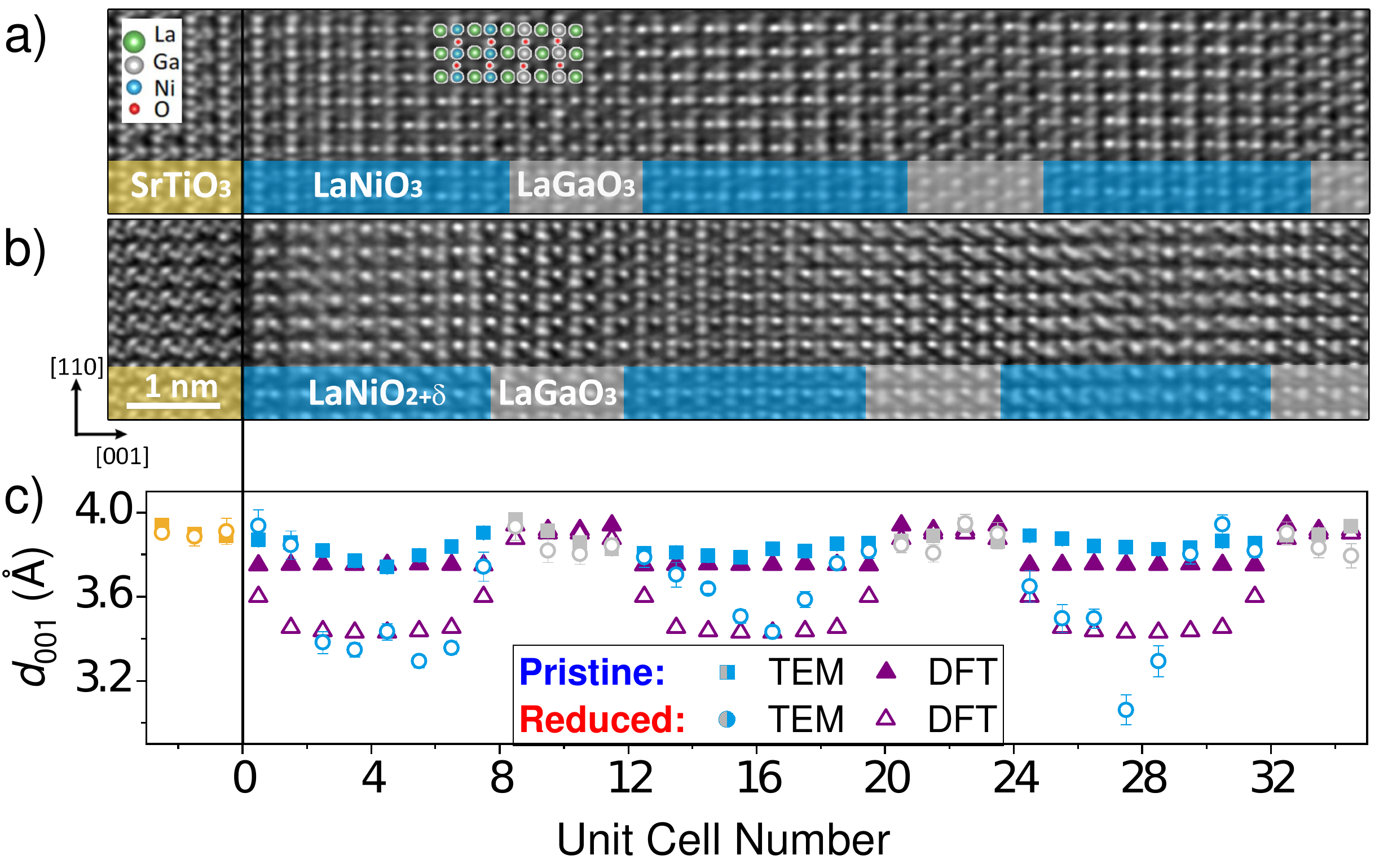}
\caption{Negative-C$_s$ HRTEM images of the $m=8, n=4$ (a) pristine and (b) reduced superlattices measured in $[1-10]_{pc}$ projection. The interfaces between SrTiO$_3$ substrate (yellow) and LaNiO$_{3\,(2+\delta)}$ (blue)-LaGaO$_3$ (grey) layer stacks are clearly shifted in the reduced sample due to the $d_{001}$ contraction upon oxygen removal. Schematic atomic model is overlaid on the HRTEM image in panel (a). (c) Values of the $d_{001}$-spacing in pristine and reduced superlattices measured from the HRTEM images (a,b), where each data point represents the average value of 15 unit cells along the $[110]_{pc}$ axis. For comparison we show the corresponding values from the DFT structure relaxation. \label{TEMIm}}
\end{figure}

\begin{figure*}[t]
  \includegraphics[width=2.0\columnwidth]{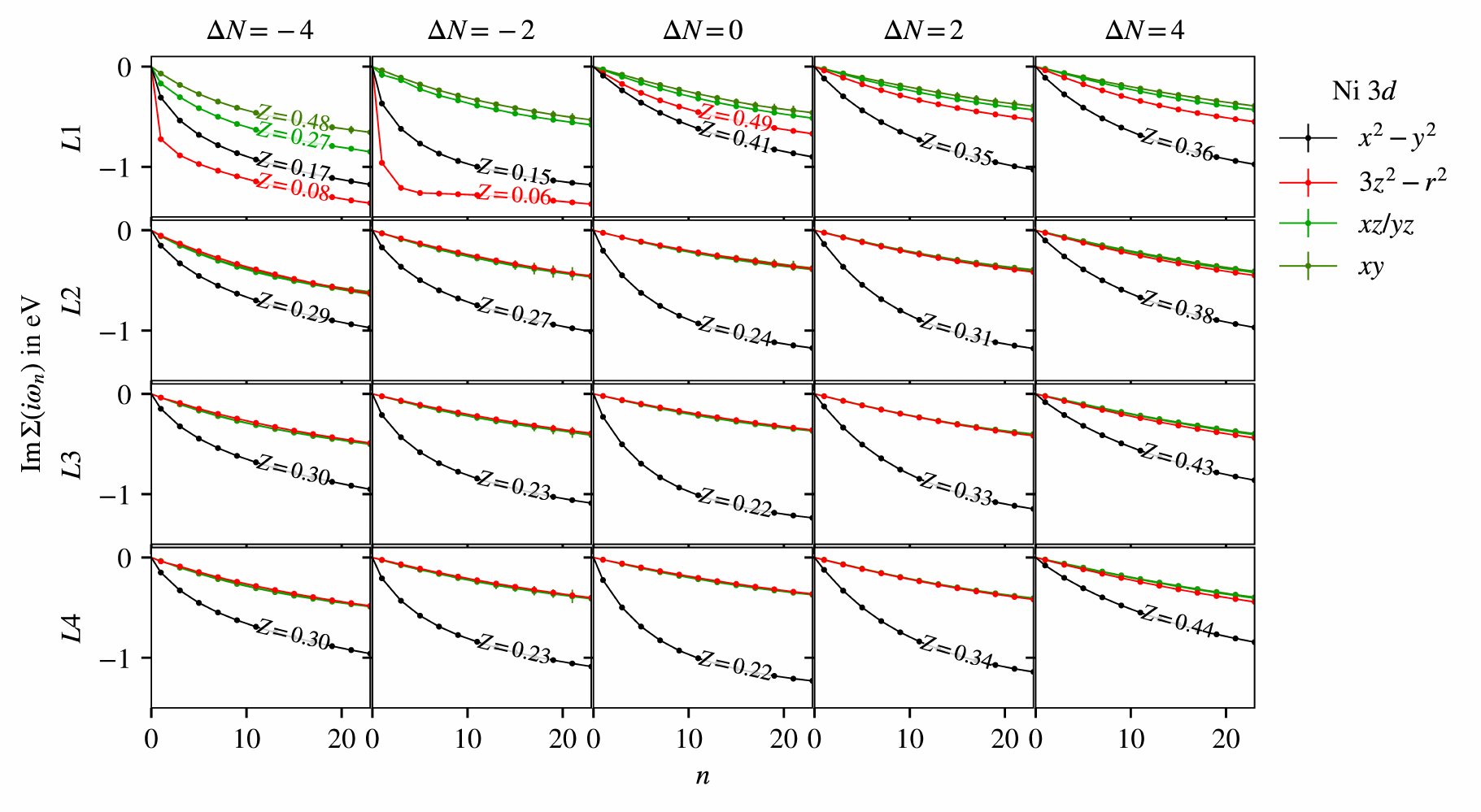}
  \caption{Layer- and orbital-resolved imaginary part of the self-energy. The lines are annotated with the corresponding quasi-particle renormalization factor where there is a quasi-particle peak present in Fig.~\ref{fig:DMFT_spectra}.\label{fig:DMFT_self_energies}}
\end{figure*}

Cross-sectional HRTEM specimens of pristine and reduced samples from the same growth (Fig.~\ref{TEM}(a,b)) were prepared by focused ion beam (FIB) technique using a NVision $040$ Ar with a special X2-holder\cite{lechner2012} for high-quality lamella preparation. We chose pseudo-cubic $(1-10)_{pc}$ plane cuts to determine accurately the atomic column distances along the superlattice growth direction.  Imaging-side spherical aberration-corrected (AC)-FEI Titan $80$~to~$300$\,kV was used in HRTEM imaging mode operated at $300$\,kV. Negative-spherical-aberration (C$_s$) imaging conditions (C$_{s} = -15\,\mu$m) was applied to image the atomic columns by white atom contrast.\cite{Jia2003} For the quantitative analysis, the AC-HRTEM images were filtered by Wiener filter to enhance the signal to noise ratio. CalAtom\cite{Zhang2019} together with a self-written Matlab code was used to extract the lattice parameter values $d_{001}$ (Fig.~\ref{TEM}(c)) within the individual LaNiO$_{3\,(2+\delta)}$ and LaGaO$_3$ layer stacks along the superlattice growth direction. To distinguish LaNiO$_{3\,(2+\delta)}$ from the LaGaO$_3$ layer stacks, we took advantage of the different pattern that results from the $Pbnm$ distortion of LaGaO$_3$ in the selected projection.\cite{Qi2015} Together with the information on the stacking sequence from the growth (starting with 8 unit cells of LaNiO$_{3\,(2+\delta)}$ from the substrate, followed by 4 unit cells of LaGaO$_3$ and so on) we identified the interfaces. Within the experimental accuracies of the methods, we ?nd good agreement between TEM and XRD data and the corresponding DFT $d_{001}$ values. The latter were derived by relaxing the ionic positions within the superlattice unit cell, whereby the in-plane lattice parameter and the bilayer thickness were fixed to the experimental XRD values of pristine and reduced superlattices, respectively.

\section{Hubbard-Kanamori Hamiltonian}
\label{app:DMFT_Kanamori}
The local interaction Hamiltonian in Hubbard-Kanamori form reads
\begin{subequations}
  \begin{align}
    H_{\mathrm{int}}
    \label{eq:Kanamori_intra}
    &= U \sum_m \hat{n}_{m, \uparrow} \hat{n}_{m, \downarrow} \\
    \label{eq:Kanamori_inter1}
    &\quad + U' \sum_{m \neq m'} \hat{n}_{m', \uparrow} \hat{n}_{m', \downarrow} \\
    \label{eq:Kanamori_inter2}
    &\quad + (U' - J_H) \sum_{m < m', \sigma} \hat{n}_{m, \sigma} \hat{n}_{m', \sigma} \\
    \label{eq:Kanamori_spinflip}
    &\quad + J_H \sum_{m \neq m'} \hat{d}_{m, \uparrow}^\dagger \hat{d}_{m', \downarrow}^\dagger \hat{d}_{m, \downarrow} d_{m', \uparrow} \\
    \label{eq:Kanamori_pairhop}
    &\quad + J_H \sum_{m \neq m'} \hat{d}_{m, \uparrow}^\dagger \hat{d}_{m, \downarrow}^\dagger \hat{d}_{m', \downarrow} d_{m', \uparrow},
  \end{align}
\end{subequations}
where \eqref{eq:Kanamori_intra} is the intra-orbital Coulomb interaction, \eqref{eq:Kanamori_inter1} and \eqref{eq:Kanamori_inter2} is the inter-orbital Coulomb interaction, \eqref{eq:Kanamori_spinflip} is the Hund's rule spin-flip interaction, and \eqref{eq:Kanamori_pairhop} is the Hund's rule pair-hopping interaction.

\section{DMFT self energies \label{selfenergies}}
\label{app:DMFT_sigma}

In Fig.~\ref{fig:DMFT_self_energies} we show the imaginary part of the layer and orbital resolved self-energy from which we can extract the QP renormalization factor for the spectra in Fig.~\ref{fig:DMFT_spectra}. The mass renormalization factor is defined as
\begin{equation}
  Z_\alpha\equiv m^\text{QP}_\alpha/m^\text{free}_\alpha\approx \frac{1}{1-\operatorname{Im}\Sigma^\text{DMFT}_\alpha(i\omega_0)/\omega_0}\, ,
\end{equation}
where $\omega_0=\pi k_B T$ is the first fermionic Matsubara frequency defined by the Boltzmann constant $k_B$ and Temperature $T$, and $\Sigma^\text{DMFT}_\alpha$ is the local DMFT self-energy for orbital $\alpha$. We only provide renormalization factors for those orbitals which contribute to the coherent QP peak at the Fermi energy.

\section{DFT and DMFT densities \label{densities}}

\begin{table}[hb]
  \caption{Layer-resolved projected DFT densities of the Ni $3d$ orbitals (per spin).\label{tab:occupations_DFT}}
  \begin{ruledtabular}
    \begin{tabular}{@{}*6c@{}}
      & $x^2-y^2$ & $3z^2-\mathbf{r}^2$ & $xz/yz$ & $xy$ & $n_{\mathrm{tot}}$ \\
      \hline
      $L1$ & 0.57 & 0.81 & 0.97 & 0.98 & 4.31 \\
      $L2$ & 0.59 & 0.83 & 0.97 & 0.98 & 4.35 \\
      $L3$ & 0.61 & 0.84 & 0.96 & 0.98 & 4.37 \\
      $L4$ & 0.61 & 0.84 & 0.96 & 0.98 & 4.37 \\
    \end{tabular}
  \end{ruledtabular}
\end{table}

\begin{table}[hb]
  \caption{DMFT densities (per spin)\label{tab:occupations_DMFT}}
  \begin{ruledtabular}
    \begin{tabular}{@{}*6c@{}}
      & $x^2-y^2$ & $3z^2-\mathbf{r}^2$ & $xz/yz$ & $xy$ & $n_{\mathrm{tot}}$ \\
      \hline
      $L1$ & 0.60 & 0.78 & 0.98 & 0.99 & 4.34 \\
      $L2$ & 0.57 & 0.86 & 0.98 & 0.99 & 4.39 \\
      $L3$ & 0.58 & 0.87 & 0.97 & 0.99 & 4.40 \\
      $L4$ & 0.58 & 0.87 & 0.97 & 0.98 & 4.40 \\
    \end{tabular}
  \end{ruledtabular}
\end{table}

\end{document}